\documentclass[11pt]{article}
\usepackage[margin=1in]{geometry}
\usepackage[T1]{fontenc}
\usepackage[nomath]{lmodern}
\usepackage{amsmath,amssymb,amsthm,mathtools,microtype,xurl,needspace,listings}
\usepackage{graphicx,tikz}
\usepackage{booktabs,tabularx}
\usetikzlibrary{arrows.meta}
\usepackage{hyperref}
\hypersetup{
  hypertexnames=false,
  colorlinks=true,
  linkcolor=black,
  citecolor=black,
  urlcolor=black,
  filecolor=black,
  pdftitle={The Dry Ten Martini Problem},
  pdfauthor={Dan S. Borgnia and Matthew Faust}
}
\newtheorem{theorem}{Theorem}[section]
\newtheorem{lemma}[theorem]{Lemma}
\newtheorem{corollary}[theorem]{Corollary}
\theoremstyle{definition}
\newtheorem{definition}[theorem]{Definition}
\newtheorem{remark}[theorem]{Remark}
\theoremstyle{plain}
\numberwithin{equation}{section}
\newcommand{\ii}{\mathrm{i}}

\newcommand{\norm}[1]{\lVert#1\rVert}

\newcommand{\VV}{\mathcal{V}}
\newcommand{\ls}{\ensuremath{L^{*}}}
\newcommand{\lext}{\ensuremath{LE}}
\newcommand{\lw}{\ensuremath{W}}
\title{The Dry Ten Martini Problem at Criticality}
\author{Dan S. Borgnia\thanks{Department of Physics, University of California, Berkeley, California 94720, USA. Email: \href{mailto:dborgnia@berkeley.edu}{\nolinkurl{dborgnia@berkeley.edu}}.}
\and Matthew Faust\thanks{Department of Mathematics, Michigan State University, East Lansing, Michigan 48824, USA; Simons Institute for the Theory of Computing, University of California, Berkeley, California 94720, USA. Email: \href{mailto:mfaust@msu.edu}{\nolinkurl{mfaust@msu.edu}}.}}
\date{}
\lstdefinestyle{lean}{basicstyle=\footnotesize\ttfamily,
  columns=fullflexible,keepspaces=true,showstringspaces=false,
  breaklines=true,breakatwhitespace=true,mathescape=true}
\begin{document}
\maketitle
\begin{center}
\itshape In memory of Matthew S. Brennan, 1994--2021
\end{center}
\begin{abstract}
We solve the dry ten martini problem (\href{https://mattfaust.github.io/Dry10/dtm_LEAN.zip}{\emph{lean-verified}}
this time) by showing that every labeled almost Mathieu gap is open for $\lambda\neq 0$. In particular, we prove that the maximum logarithmic height of each labeled gap is non-decreasing in the coupling $\lambda$ and extend an initial open \emph{seed gap} for each gap label $n$ from $\lambda\in(0,\lambda_{n})$ to all $\lambda > 0$. The non-decreasing property follows from restoring time-reversal symmetry to the almost Mathieu operator $H$ by pairing it with its magnetic inverse $\bar{H}$. This allows us to express the logarithmic coupling derivative of the maximal gap height as the expectation value of a non-negative operator, yielding the required monotonicity.
\end{abstract}

{\small\noindent
\textbf{Lean labels.} \ls: same proof; \lext: cited inputs replaced by checked counterparts;
\lw: unverified, unused background.
Dependencies retain their own status. Appendix~\ref{app:index} gives the proof correspondence.
\par}

\section{Introduction}
\label{sec:intro}
We consider the almost Mathieu operator (AMO) $h_{\lambda, \alpha, \varphi} \in
\mathcal{B}(\ell^{2}(\mathbb{Z}))$,
with coupling $\lambda \in \mathbb{R}\setminus\{0\}$, irrational frequency $\alpha$,
and phase $\varphi \in \mathbb{R}/(2\pi\mathbb{Z})$, given by
\begin{align}
  (h_{\lambda, \alpha, \varphi}\psi)_{j}
  &= \psi_{j+1}+\psi_{j-1}+2\lambda\cos(2\pi\alpha j+\varphi)\psi_{j}.
  \label{eq:amo}
\end{align}
Its spectrum $\Sigma_{\lambda, \alpha}$ is phase independent \cite{AS}.
The bounded components of its complement are the \emph{open gaps}.
From now on, $\alpha$ is fixed and $\Sigma_{\lambda} = \Sigma_{\lambda, \alpha}$.
Write $h_{\lambda, \varphi} = h_{\lambda, \alpha, \varphi}$, and let $(e_{j})_{j \in
\mathbb{Z}}$
be the canonical basis of $\ell^{2}(\mathbb{Z})$.

Harper introduced the model for conduction electrons in a magnetic field
\cite{Harper}; it later became a lattice model of the integer quantum Hall
effect \cite{TKNN}. Spinless electrons hop between neighboring sites of
$\mathbb{Z}^{2}$ with flux $\Theta = 2\pi\alpha$ through each plaquette.
On $\ell^{2}(\mathbb{Z}^{2})$, with canonical basis $\tilde{e}_{n, m}$,
the Landau-gauge magnetic translations are
\begin{align}
  U\tilde{e}_{n, m} &= \tilde{e}_{n-1, m}, \qquad
  V\tilde{e}_{n, m} = e^{\mathrm{i} \Theta n}\tilde{e}_{n, m-1}.
  \label{eq:magnetictranslations}
\end{align}
They satisfy $UV = e^{\mathrm{i} \Theta}VU$. We define the \emph{magnetic rotation
algebra} to be the
unital $C^{*}$-algebra generated by $U, V$ \cite[p.~417]{Rieffel}:
\begin{align}
  \mathcal{A}_{\alpha} &:= C^{*}(U, V)
  = \overline{\operatorname{span}\{U^{p}V^{q}:p, q \in \mathbb{Z}\}}^{\lVert\cdot\rVert}
  \subset \mathcal{B}(\ell^{2}(\mathbb{Z}^{2})).
  \label{eq:rotation-algebra}
\end{align}
Write $e_{0, 0} := \tilde{e}_{0, 0}$ for the origin.
The Harper operator and its resolvent are
\begin{align}
  H_{\lambda} &= U+U^{*}+\lambda(V+V^{*}), \qquad
  R_{\lambda}(z) := (z-H_{\lambda})^{-1}\quad(z \notin \Sigma_{\lambda}).
  \label{eq:harper}
\end{align}
Partial Fourier transformation in the second coordinate is the unitary map
$(\mathcal{F}_{y}\tilde{e}_{n, m})(\varphi) = e^{-\mathrm{i} m \varphi}e_{n}$, with measure
$\mathrm{d} \varphi/(2\pi)$, decomposing $H_{\lambda}$ into the fibers
\eqref{eq:amo}, so $\sigma(H_{\lambda}) = \Sigma_{\lambda}$. For $A \in \mathcal{A}_{\alpha}$
with fibers $A(\varphi)$, the \emph{canonical trace} is
\begin{align}
  \tau_{\alpha}(A) &:= \langle e_{0, 0}, Ae_{0, 0}\rangle
  = \int_{0}^{2\pi}\langle e_{0}, A(\varphi)e_{0}\rangle\frac{\mathrm{d} \varphi}{2\pi}.
  \label{eq:canonical-trace}
\end{align}
It is positive and normalized, with $\tau_{\alpha}(AB) = \tau_{\alpha}(BA)$ for
$A, B \in \mathcal{A}_{\alpha}$ \cite[Secs.~3.1 and~3.6]{BES}. We use
\eqref{eq:canonical-trace} to write the fiber integrals in trace notation.
Fourier transforming the other coordinate, or making a magnetic quarter-turn
in two dimensions, gives Aubry--Andr\'e duality \cite{AA}:
\begin{align}
  \mathcal{R}\tilde{e}_{n, m} &= e^{\mathrm{i} \Theta nm}\tilde{e}_{-m, n}, \qquad
  \mathcal{R} H_{\lambda}\mathcal{R}^{*} = \lambda H_{1/\lambda}.
  \label{eq:quarterturn}
\end{align}
The sign involution is
\begin{align}
  \mathcal{S}\tilde{e}_{n, m} &= (-1)^{m}\tilde{e}_{n, m}, \qquad
  \mathcal{S} H_{\lambda}\mathcal{S}^{*} = H_{-\lambda}.
  \label{eq:signunitary}
\end{align}
Both unitaries fix $e_{0, 0}$. We write $k_{y} := \varphi$ for the Fourier variable dual to the
second lattice coordinate, $y$, and $k_{x}$ for the Fourier variable dual to the
first, $x$.

\subsection{Fermi projectors and gap labels}
For Harper's noninteracting, particle-number-conserving model,
an initial Slater-determinant state evolves by propagating its occupied
single-particle orbitals under the action of the one-particle magnetic
Hamiltonian \cite[Sec.~1]{DiVincenzoTerhal2005}.
The zero-temperature single-particle occupied subspace is given by the Fermi projector:
\begin{definition}[Fermi projector and density of states]
\label{def:fermi}
For real $E \notin \Sigma_{\lambda}$, choose a counterclockwise contour
$\Gamma_{\lambda, E} \subset \mathbb{C}\setminus\Sigma_{\lambda}$ enclosing precisely
the spectrum
below $E$. The Fermi projector is given by the Riesz formula,
\begin{align}
  \chi_{H_{\lambda}}(E) &:= \frac{1}{2\pi\mathrm{i}}\oint_{\Gamma_{\lambda, E}}R_{\lambda}(z)\,\mathrm{d} z
  \in \mathcal{A}_{\alpha}.
  \label{eq:fermi}
\end{align}
Spectral calculus identifies this with the projection onto energies at most $E$. For any
bounded self-adjoint $A$, the Borel formula
\begin{align}
  \chi_{A}(E) &:= \mathbf{1}_{(-\infty, E]}(A),
  \label{eq:fermidefinition}
\end{align}
with $\mathbf{1}_{S}$ the indicator function on the set $S$, extends the definition to
all real $E$.
The origin pairings of these projectors define the integrated density of states and
associated probability measure $\nu_{\lambda}$:
\begin{align}
  N_{\lambda}(E) &:= \langle e_{0, 0}, \chi_{H_{\lambda}}(E)e_{0, 0}\rangle,
  \qquad \nu_{\lambda}(({-\infty}, E]) = N_{\lambda}(E).
  \label{eq:dos}
\end{align}
\end{definition}
\begin{remark}
The projector construction \eqref{eq:fermi}--\eqref{eq:dos} and
\eqref{eq:canonical-trace} give $N_{\lambda}(E) = \tau_{\alpha}(\chi_{H_{\lambda}}(E))$
in a gap, and spectral calculus gives
$\int f\,\mathrm{d} \nu_{\lambda} = \tau_{\alpha}(f(H_{\lambda}))$ for continuous $f$.
At spectral energies, \eqref{eq:dos} uses the origin pairing of the Borel
projector, which need not belong to $\mathcal{A}_{\alpha}$.
\end{remark}
Writing $\{x\} := x-\lfloor x\rfloor$ for the fractional part, we relate
$N_{\lambda}(E)$ to $\alpha$:
\begin{theorem}[Gap labeling \cite{GapLabel}; \lw]
\label{thm:gaplabel}
On each bounded gap, $N_{\lambda}(E) = \{n\alpha\}$ for a unique
$n \in \mathbb{Z}\setminus\{0\}$.
\end{theorem}

\begin{proof}[Proof sketch]
The Riesz projector belongs to $\mathcal{A}_{\alpha}$ by \eqref{eq:fermi},
and its trace lies in $(\mathbb{Z}+\alpha\mathbb{Z})\cap[0, 1]$ \cite{PV,Rieffel}.
The projector is constant on a gap; full support of $\nu_{\lambda}$ gives
distinct IDS values on distinct gaps \cite{AS}, and
$\alpha \notin \mathbb{Q}$ makes the labels unique.
\end{proof}

With compatible orientation conventions, the integer $n$ is the Chern
number governing the quantized Hall conductance \cite{TKNN,BES}.
Bulk--edge correspondence relates it to boundary spectral flow and
quantized edge conductance
\cite{ProdanSchulzBaldes2016,KellendonkRichterSchulzBaldes2002}, modelling the integer
quantum Hall effect.

\subsection{The Ten Martini problems}
In 1981, Kac famously asked whether ``all gaps are there'' and offered
ten martinis for a solution \cite[Remark~1.1(e)]{LDZCMV}.
Simon published the challenge \cite{Simon}, leading to decades of
work on AMO spectrum. The question came to be studied in two forms:
the Ten Martini problem asks for Cantor spectrum, so that open gaps accumulate at every spectral energy; the \emph{dry} Ten Martini problem retains the full gap-opening requirement, that is, every label in Theorem~\ref{thm:gaplabel} must be attained by an open gap.

In a milestone result, Avila and Jitomirskaya solved the Ten Martini
problem \cite{AJ} and showed that open gaps accumulate at every
spectral energy:
\begin{theorem}[Ten Martini; Avila--Jitomirskaya; \lw]
\label{thm:cantor}
For every irrational frequency and every nonzero coupling, $\Sigma_{\lambda}$
is a Cantor set.
\end{theorem}
Reducibility, localization, and cocycle methods gave successive
gap-opening results \cite{Eliasson,Puig,AJAL,AvilaGlobal};
Avila--You--Zhou proved the dry theorem for every noncritical AMO
coupling \cite{AYZ}. Building the Type~I framework
\cite{GJY_type}, Ge--Jitomirskaya--You established that analytic Type~I one-frequency
quasiperiodic Schr\"odinger
operators have Cantor spectrum for every irrational frequency, with
stability in the appropriate analytic topology \cite{GJY2026}.

For the stronger dry statement, Li \cite{Li2026} and Ge \cite{Ge2026}
gave different proofs that, for every irrational frequency and real
analytic potential, every subcritical Type~I spectral energy satisfying
the gap-labelling condition is an endpoint of an open gap. Ge's proof
supplies the details of the argument announced in
\cite[Remark~1.3]{GJY2026}. In the supercritical regime, Li--Xu--Zhou proved
the corresponding Type~I dry statement and, in particular, stability of
all supercritical AMO gaps under sufficiently small analytic
perturbations \cite{LXZStability}.

At critical coupling, Choi--Elliott--Yui proved full gap opening for a subclass of Liouville frequencies \cite{CEY}.
Figueras and Puig obtained rigorous computer-assisted proofs of individual labeled gaps at specific irrational frequencies, together with bounds on their widths \cite{FiguerasPuig2026}. However, the full dry Ten Martini problem for arbitrary irrational frequency at
critical coupling remained open.
Although this constitutes a single coupling value $\lambda =1$, it marks a change of spectral type: for
Diophantine frequencies and almost every phase, the spectrum is purely absolutely
continuous for $0 < \lambda < 1$, pure point for $\lambda > 1$, and purely singular
continuous at $\lambda = 1$ \cite{Jit99,JitCritical}. Moreover, singular continuous
spectra are Baire generic in suitable families of self-adjoint operators
\cite{SimonGeneric}, suggesting the failure of existing methods at $\lambda =1$ constituted a fundamental obstruction.

\begin{figure}[htbp]
    \centering
    \includegraphics[width=\linewidth]{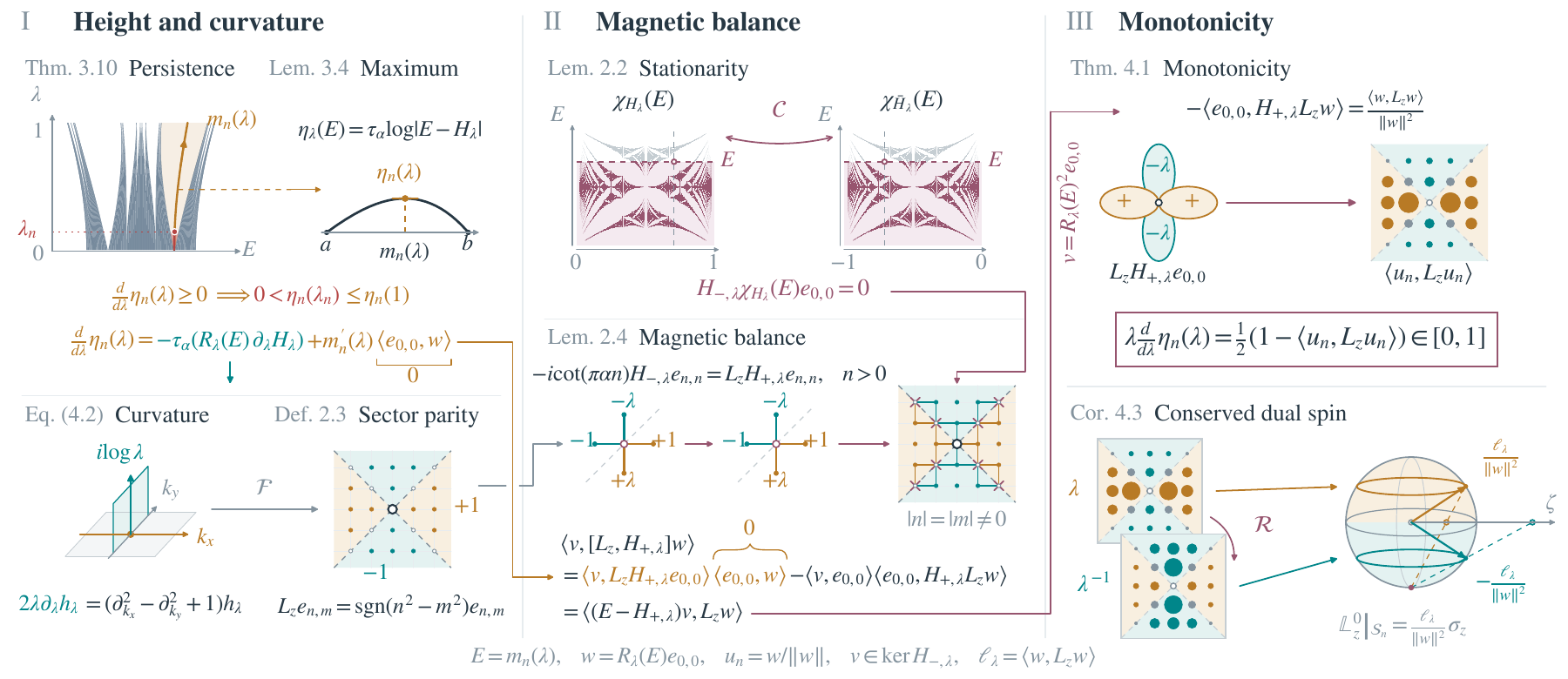}
    \caption{\textbf{I.} Selecting the gap maximum removes the energy term in the total
    derivative (gold). The remaining term (teal) is expressed via the sector parity
    $L_{z}$. \textbf{II.} Magnetic inversion pairs the Fermi-projector sources at the
    origin, giving stationarity. This cancels nonzero diagonal pairings, and magnetic
    balance leaves the origin terms. \textbf{III.} Choosing $v = R_{\lambda}(E)^{2}e_{0,
    0}$ identifies the surviving origin response with the normalized $L_{z}$, yielding
    the required  $0 \leq \lambda\eta_{n}'(\lambda) \leq 1$ to keep the (red) seed gap
    open up to $\lambda = 1$. Below, quarter-turn duality gives the conserved spin
    doublet. References identify the proof steps, arrows indicate dependencies, and the
    coupling spectrum uses $\alpha\simeq\sqrt{2}/2$.}
    \label{fig:proof_narrative}
\end{figure}

In this manuscript, motivated by the extensive literature on invertible phases of matter
\cite{Wen2017,GaiottoJohnsonFreyd2019}, we pair the AMO with its time-reversed magnetic
action to restore time-reversal symmetry on the spectral source space\footnote{The same time-reversal symmetric pairing was used to study phase-averaged transport for the critical AMO \cite{bellissard2002phase} and in multiple works on quasiperiodic operators by Riedel~\cite{Riedel,riedel_1988}.}. We use the
resulting magnetic balance identity to arrive at the bound in
Theorem~\ref{thm:response}, \emph{closing the gap}\footnote{Phrasing proposed by Wencai
Liu.} and proving that all gaps are open\footnote{See Fig~\ref{fig:proof_narrative} for
an annotated sketch of the full argument.}:
\begin{theorem}[Dry Ten Martini; \lext]
\label{thm:main}
For every irrational $\alpha$, every $\lambda \neq 0$, and every $n \in
\mathbb{Z}\setminus\{0\}$, there is a bounded open gap on which $N_{\lambda}(E) =
\{n\alpha\}$.
\end{theorem}
The noncritical case is due to Avila--You--Zhou \cite{AYZ}. However, we highlight the
scope of the balance identity by treating all nonzero couplings directly. Indeed, the
magnetic balance refines Aubry--Andr\'e duality~\cite{AA} into a conserved
$\mathrm{SU}(2)$ doublet for the centered dual pair (Corollary~\ref{cor:tangentspin}).

\section{Restoring Time-Reversal Symmetry}
\label{sec:time_reversal}
Reversing the magnetic phases conjugates the hopping amplitudes. We implement this in
the symmetric gauge, starting from the Landau basis
\eqref{eq:magnetictranslations}:
\begin{align}
  e_{n, m} &:= e^{-\mathrm{i} \pi\alpha nm}\tilde{e}_{n, m}, \qquad
  (\mathcal{F}_{y} e_{n, m})(k_{y}) = e^{-\mathrm{i} \pi\alpha nm}e^{-\mathrm{i} m k_{y}}e_{n}.
  \label{eq:basis}
\end{align}
Define the \emph{magnetic inversion} $\mathcal{C}$ to be coefficientwise complex
conjugation in this basis, so $\mathcal{C}e_{n, m} = e_{n, m}$. In the Landau basis, for
$v \in \ell^{2}(\mathbb{Z}^{2})$, this reads
\begin{align}
  \langle\tilde{e}_{n, m}, \mathcal{C}v\rangle
  &= e^{-\mathrm{i} \Theta nm}\overline{\langle\tilde{e}_{n, m}, v\rangle}.
  \label{eq:conjugation}
\end{align}
Thus $\mathcal{C}$ is an anti-unitary involution, with
$\mathcal{C}^{2} = I$, $\mathcal{C}^{*} = \mathcal{C}$, and
$\mathcal{C}e_{0, 0} = e_{0, 0}$.

\begin{definition}[Magnetic inverse]
\label{def:magneticinverse}
The \emph{magnetic-inverse Hamiltonian} is
\begin{align}
  \bar{H}_{\lambda} &:= \mathcal{C}H_{\lambda} \mathcal{C}^{*} \in \mathcal{C}\mathcal{A}_{\alpha} \mathcal{C}^{*} .
\end{align}
The algebra $\mathcal{C}\mathcal{A}_{\alpha}\mathcal{C}^{*}$ has opposite flux, since
\begin{align}
  (\mathcal{C}U\mathcal{C}^{*})(\mathcal{C}V\mathcal{C}^{*}) &= e^{-\mathrm{i} \Theta}(\mathcal{C}V\mathcal{C}^{*})(\mathcal{C}U\mathcal{C}^{*}).
  \label{eq:reversedflux}
\end{align}
\end{definition}
For $t \in \mathbb{R}$, anti-unitarity reverses time and exchanges the
Fermi projectors:
\begin{align}
  \mathcal{C}e^{-\mathrm{i} tH_{\lambda}}\mathcal{C}^{*} &= e^{\mathrm{i} t\bar{H}_{\lambda}}, \qquad
  \mathcal{C}\chi_{H_{\lambda}}(E)\mathcal{C}^{*} = \chi_{\bar{H}_{\lambda}}(E).
  \label{eq:timereversal}
\end{align}
Let $\mathcal{V}_{\lambda}$ be the closed span of
$\chi_{H_{\lambda}}(E)e_{0, 0}$ over real $E$,
\begin{lemma}[Stationarity; \ls]
\label{lem:stationarity}
The operators $H_{\lambda}$ and $\bar{H}_{\lambda}$ leave
$\mathcal{V}_{\lambda}$ invariant and coincide on it. In particular,
for every $E \in \mathbb{R}$,
\begin{align}
  (H_{\lambda}-\bar{H}_{\lambda})
  \chi_{H_{\lambda}}(E)e_{0, 0} &= 0.
  \label{eq:vectorizedstationarity}
\end{align}
\end{lemma}

\begin{proof}
In the symmetric basis,
\begin{align}
  Ue_{n, m} &= e^{-\mathrm{i} \pi\alpha m}e_{n-1, m}
  , \qquad
  Ve_{n, m} = e^{\mathrm{i} \pi\alpha n}e_{n, m-1}.
\end{align}
So, the reflection $e_{n, m} \mapsto e_{n, -m}$
sends $U$ to $\mathcal{C} U\mathcal{C}^{*}$ and $V$ to
$\mathcal{C} V^{*}\mathcal{C}^{*}$, and hence sends
$H_{\lambda}$ to $\bar{H}_{\lambda}$.

Writing $[A, B] := AB-BA$, the original and reflected shifts satisfy
$[U, \mathcal{C} U\mathcal{C}^{*}] = [V, \mathcal{C} V\mathcal{C}^{*}] = 0$,
and by direct computation,
\begin{align}
  U(\mathcal{C} V\mathcal{C}^{*})e_{n, m}
  &= (\mathcal{C} V\mathcal{C}^{*})Ue_{n, m}
  = e^{-\mathrm{i} \pi\alpha(n+m-1)}e_{n-1, m-1}, \notag \\
  V(\mathcal{C} U\mathcal{C}^{*})e_{n, m}
  &= (\mathcal{C} U\mathcal{C}^{*})Ve_{n, m}
  = e^{\mathrm{i}\pi\alpha(n+m-1)}e_{n-1, m-1}.
  \label{eq:commutation}
\end{align}
By unitarity, the same is true for adjoints, and at the origin,
$\exp(\mathrm{i}\pi\alpha 0) = 1$. So,
\begin{align}
  [H_{\lambda}, \bar{H}_{\lambda}] &= 0\qquad\text{and}\qquad
  H_{\lambda} e_{0, 0} = \bar{H}_{\lambda} e_{0, 0}.
  \label{eq:commonorigin}
\end{align}
This implies commutation of $\bar{H}_{\lambda}$ with continuous functions of $H_{\lambda}$ by polynomial approximation,
and with spectral projections of $H_{\lambda}$ by strong limits, i. e., $\chi_{H_\lambda}(E)$ commutes with $\bar H_\lambda$. Hence,
\begin{align}
  (H_{\lambda}-\bar{H}_{\lambda})
  \chi_{H_{\lambda}}(E)e_{0, 0}
  &= \chi_{H_{\lambda}}(E)
  (H_{\lambda}-\bar{H}_{\lambda})e_{0, 0} = 0.
\end{align}
where the last equality follows from the second part of \eqref{eq:commonorigin}. The
bounded operator $H_{\lambda}-\bar{H}_{\lambda}$ therefore vanishes on every generator
of $\mathcal{V}_{\lambda}$, hence on their closed span. This space is invariant under
$H_{\lambda}$ by spectral calculus, and therefore also
under $\bar{H}_{\lambda}$.
\end{proof}

Define $H_{ \pm , \lambda} = \frac{1}{2}(H_{\lambda} \pm \bar{H}_{\lambda})$.
The symmetrized Hamiltonian is invariant under $\mathcal{C}$ by construction:
\begin{align}
  \mathcal{C}H_{+, \lambda}\mathcal{C}^{*} &= H_{+, \lambda}.
\end{align}
Stationarity makes this symmetrization exact on the spectral source space:
\begin{align}
  H_{-, \lambda}|_{\mathcal{V}_{\lambda}} &= 0, \qquad H_{+, \lambda}|_{\mathcal{V}_{\lambda}} = H_{\lambda}|_{\mathcal{V}_{\lambda}},
  \qquad \sigma(H_{+, \lambda}|_{\mathcal{V}_{\lambda}}) = \Sigma_{\lambda}.
  \label{eq:stationarity}
\end{align}
The spectral equality follows because the origin measure is
$\nu_{\lambda}$, whose support is $\Sigma_{\lambda}$ \cite{AS}.
This space is also $\mathcal{C}$-invariant. Indeed, $\mathcal{C}$ fixes $e_{0, 0}$,
\begin{align}
  \mathcal{C}\chi_{H_{\lambda}}(E)e_{0, 0}
  &= \chi_{\bar{H}_{\lambda}}(E)e_{0, 0}
  = \chi_{H_{\lambda}}(E)e_{0, 0}.
  \label{eq:sourceconjugation}
\end{align}
Consequently, the dynamics generated by $H_{\lambda}$ on $\mathcal{V}_{\lambda}$ satisfy
\begin{align}
  \mathcal{C}e^{-\mathrm{i} tH_{\lambda}}v &= e^{\mathrm{i} tH_{\lambda}}\mathcal{C}v,
  \qquad v \in \mathcal{V}_{\lambda}, \quad t \in \mathbb{R}.
  \label{eq:restrictedtimereversal}
\end{align}
This is time-reversal symmetry on the spectral source space.

In the symmetric basis, the even and odd combinations under $\mathcal{C}$ act by
\begin{align}
  2H_{ \pm , \lambda}e_{n, m}
  &= \sum_{\rho = \pm 1}\Bigl[
  (e^{\mathrm{i} \rho\pi\alpha m} \pm e^{-\mathrm{i} \rho\pi\alpha m})e_{n+\rho, m}
  +\lambda(e^{-\mathrm{i} \rho\pi\alpha n} \pm e^{\mathrm{i} \rho\pi\alpha n})e_{n, m+\rho}\Bigr].
  \label{eq:latticeactions}
\end{align}
The quarter-turn satisfies $\mathcal{R} H_{ \pm , \lambda}\mathcal{R}^{*} = \lambda H_{
\pm , 1/\lambda}$. We introduce an observable that changes sign when the two hopping
directions are exchanged:
\begin{definition}[Sector parity]
\label{def:magneticsectors}
Define
\begin{align}
  L_{z}e_{n, m} &= \operatorname{sgn}(n^{2}-m^{2})e_{n, m}, \qquad \operatorname{sgn}(0) = 0.
  \label{eq:angular}
\end{align}
Under Aubry duality, $\mathcal{R} L_{z}\mathcal{R}^{*} = -L_{z}$, and under
time reversal, $\mathcal{C}L_{z}\mathcal{C}^{*} = L_{z}$.
\end{definition}
The diagonals $m = \pm n$ divide the lattice into four sectors.
We call the two regions with $|n| > |m|$ the horizontal sectors
and those with $|m| > |n|$ the vertical sectors.
Thus $L_{z}$ equals $+1$ on the horizontal sectors, $-1$ on the
vertical sectors, and $0$ on the separating diagonals. Its commutator
with the hopping operator is supported on bonds meeting those diagonals.

Stationarity cancels the nonzero diagonal contributions in the following
commutator pairings, leaving only the origin.
\begin{lemma}[Magnetic balance; \ls]
\label{lem:balance}
For $v, w \in \ker H_{-, \lambda}$,
\begin{align}
  \langle v, [L_{z}, H_{+, \lambda}]w\rangle
  &= \langle v, L_{z}H_{+, \lambda}e_{0, 0}\rangle\langle e_{0, 0}, w\rangle -\langle v, e_{0, 0}\rangle\langle L_{z}H_{+, \lambda}e_{0, 0}, w\rangle.
  \label{eq:sourcecurrent}
\end{align}
\end{lemma}
\begin{proof}
Let $\Pi_{d}$ project onto the separating diagonals $|n| = |m|$.
Every bond that changes sector touches a diagonal, so
\begin{align}
  [L_{z}, H_{+, \lambda}] &= L_{z}H_{+, \lambda}\Pi_{d}-\Pi_{d}H_{+, \lambda}L_{z}.
  \label{eq:diagonalsupport}
\end{align}
For $n \neq 0$, applying $L_{z}$ to the diagonal hops in
\eqref{eq:latticeactions} and using \eqref{eq:angular} gives
\begin{align}
  \sin(\pi\alpha n)L_{z}H_{+, \lambda}e_{n, \pm n}
  &= \mp \mathrm{i} \operatorname{sgn}(n)\cos(\pi\alpha n)H_{-, \lambda}e_{n, \pm n}.
  \label{eq:boundarybalance}
\end{align}
Since $\sin(\pi\alpha n) \neq 0$ and $v, w \in \ker H_{-, \lambda}$,
\eqref{eq:boundarybalance} gives
\begin{align}
  \langle v, L_{z}H_{+, \lambda}e_{n, \pm n}\rangle
  &= \langle w, L_{z}H_{+, \lambda}e_{n, \pm n}\rangle = 0\qquad(n \neq 0).
\end{align}
At the origin $H_{-, \lambda}e_{0, 0} = 0$, but \eqref{eq:boundarybalance} degenerates
to $0 = 0$, so it supplies no cancellation at the origin.
Expanding $\Pi_{d}$ therefore leaves only the origin:
\begin{align}
  \langle v, L_{z}H_{+, \lambda}\Pi_{d} w\rangle
  &= \langle v, L_{z}H_{+, \lambda}e_{0, 0}\rangle\langle e_{0, 0}, w\rangle, \notag\\
  \langle v, \Pi_{d} H_{+, \lambda}L_{z} w\rangle
  &= \langle v, e_{0, 0}\rangle\langle L_{z}H_{+, \lambda}e_{0, 0}, w\rangle.
  \label{eq:origin_terms}
\end{align}
Subtracting the equations in \eqref{eq:origin_terms} and substituting into
\eqref{eq:diagonalsupport} proves \eqref{eq:sourcecurrent}.
\end{proof}

\section{Proof of Main Result}
\label{sec:height}
In Theorem~\ref{thm:response}, Lemma~\ref{lem:balance} is used to establish monotonicity
of the maximum logarithmic height, defined below. We combine this with three external
inputs to prove gap persistence up to criticality: one seed gap per label~\cite{HKS},
the Thouless formula~\cite{Thouless}, and the Bourgain--Jitomirskaya formula for the
logarithmic height function on the spectrum~\cite{BJ}. In what follows, assume $\lambda > 0$.

\Needspace{6\baselineskip}
\subsection{The logarithmic height}
\begin{definition}[Logarithmic potential]
\label{def:height}
For real $E$, define
\begin{align}
  \eta_{\lambda}(E) &:= \int\log|E-t|\,\mathrm{d} \nu_{\lambda}(t).
  \label{eq:potential}
\end{align}
\end{definition}

\begin{theorem}[Thouless \cite{Thouless} and Bourgain--Jitomirskaya \cite{BJ}; \lext]
\label{thm:normalization}
The potential is finite and continuous in real energy, nonnegative, and satisfies
\begin{align}
  \eta_{\lambda}(E) &= \max\{0, \log\lambda\}, \qquad E \in \Sigma_{\lambda}.
  \label{eq:spectralvalue}
\end{align}
\end{theorem}
\begin{proof}
The Thouless formula \cite{Thouless,AS}, valid at every real energy
by \cite[Theorem~4.2]{CS}, identifies $\eta_{\lambda}(E)$ with the one-step
transfer matrix Lyapunov exponent of the associated almost
Mathieu operator.

At each fixed real energy, the transfer matrices are uniformly
bounded in phase and have determinant one. Their Lyapunov
exponent, and hence $\eta_{\lambda}(E)$, is therefore finite and
nonnegative.

Continuity in real energy follows from \cite[Theorem~1]{BJ}.
Finally, \cite[Corollary~2]{BJ}, with cosine amplitude $2\lambda$
in our notation, gives
\begin{align*}
  \eta_{\lambda}(E) &= \max\{0, \log\lambda\},
  \qquad E \in \Sigma_{\lambda}. \qedhere
\end{align*}
\end{proof}

\begin{lemma}[Upper semicontinuity; \ls]
\label{lem:usc}
The map $(\lambda, E) \mapsto \eta_{\lambda}(E)$ is upper semicontinuous
on $(0, \infty)\times\mathbb{R}$.
\end{lemma}

\begin{proof}
For $0 < \varepsilon \leq 1$, set
\begin{align}
  F_{\varepsilon}(\lambda, E)
  := \tau_{\alpha}\bigl(\log(|E-H_{\lambda}|+\varepsilon I)\bigr) = \int_{\mathbb{R}}\log(|E-t|+\varepsilon)\,\mathrm{d} \nu_{\lambda}(t).
\end{align}
Norm continuity of $H_{\lambda}$ and continuous functional calculus
show that each $F_{\varepsilon}$ is continuous in $(\lambda, E)$. As $\varepsilon$
decreases to zero, these integrals decrease, and
\begin{align}
  \lim_{\varepsilon \to 0^{+}}F_{\varepsilon}(\lambda, E)
  &= \eta_{\lambda}(E).
\end{align}
Indeed, at each fixed $(\lambda, E)$ the integrands decrease and
are uniformly bounded above on the compact support of
$\nu_{\lambda}$, so monotone convergence applies after subtraction
from a constant upper bound.
Thus,
\begin{align}
  \eta_{\lambda}(E) &= \inf_{0 < \varepsilon \leq 1}
  F_{\varepsilon}(\lambda, E),
\end{align}
showing upper semicontinuity.
\end{proof}

\subsection{The maximum logarithmic height}

\begin{lemma}[Maximum height condition; \ls]
\label{lem:maximum}
Every bounded gap of label $n$ has a unique maximizing energy
$m_{n}(\lambda)$, characterized by
\begin{align}
  \tau_{\alpha}(R_{\lambda}(m_{n}(\lambda))) &= 0.
\end{align}
It depends locally real-analytically on $\lambda$, and
$\eta_{\lambda}(m_{n}(\lambda)) > \max\{0, \log\lambda\}$.\footnote{The
characterization of maximum height by a vanishing averaged
Green function was highlighted to D. S. B. by Qi Zhou in 2024.
Riedel~\cite{Riedel} considered a related approach based on
the logarithmic potential and its critical points.}
\end{lemma}
\begin{proof}
Differentiating \eqref{eq:potential} inside a gap gives
\begin{align}
  \partial_{E}\eta_{\lambda}(E)
  &= \tau_{\alpha}(R_{\lambda}(E)) \quad \text{and}\quad
  \partial_{E}^{2}\eta_{\lambda}(E)
  = -\tau_{\alpha}(R_{\lambda}(E)^{2})
  = -\lVert R_{\lambda}(E)e_{0, 0}\rVert^{2} < 0.
  \label{eq:greenderivative}
\end{align}
The equal endpoint values from Theorem~\ref{thm:normalization} and strict
concavity give a unique interior maximum above that value. Joint
resolvent analyticity and the nonzero second derivative give local
real analyticity by the implicit-function theorem.
\end{proof}

Define the maximum height and its total derivative
\begin{align}
  \eta_{n}(\lambda) &:= \eta_{\lambda}(m_{n}(\lambda)), \qquad
  \eta_{n}'(\lambda) := \frac{\mathrm{d}}{\mathrm{d} \lambda}
  \bigl[\eta_{\lambda}(m_{n}(\lambda))\bigr].
  \label{eq:selectedheight}
\end{align}

\begin{lemma}[Maximum-height derivative; \ls]
\label{lem:heightderivative}
At every positive coupling where the bounded gap of label $n$ is open,
its maximum height satisfies
\begin{align}
  \lambda\eta_{n}'(\lambda)
  &= -\operatorname{Re}\langle R_{\lambda}(m_{n}(\lambda))e_{0, 0},
  \lambda(\partial_{\lambda} H_{+, \lambda})e_{0, 0}\rangle.
  \label{eq:maxheightinsertion}
\end{align}
\end{lemma}

\begin{proof}
Fix a gap energy $E$. The operator $(E-H_{\lambda})^{2}$ remains uniformly
positive for nearby couplings. Since
$\eta_{\lambda}(E) = \frac{1}{2}\tau_{\alpha}\!\left(
 \log\bigl((E-H_{\lambda})^{2}\bigr)\right)$,
differentiating and using trace cyclicity yields
\begin{align}
  \partial_{\lambda}\eta_{\lambda}(E)
  &= \frac{1}{2}\tau_{\alpha}\left(
  R_{\lambda}(E)^{2}\,\partial_{\lambda}[(E-H_{\lambda})^{2}]\right)
  = -\tau_{\alpha}\left(R_{\lambda}(E)\partial_{\lambda} H_{\lambda}\right).
\end{align}
At $E = m_{n}(\lambda)$, $\partial_{E}\eta_{\lambda}(E) = 0$, so the chain rule gives
\begin{align}
  \eta_{n}'(\lambda)
  &= \partial_{\lambda}\eta_{\lambda}(E)
  +m_{n}'(\lambda)\partial_{E}\eta_{\lambda}(E)
  = -\tau_{\alpha}\!\left(R_{\lambda}(E)\partial_{\lambda} H_{\lambda}\right).
\end{align}
Finally, the self-adjointness of
$R_{\lambda}(E)$ and
$(\partial_{\lambda} H_{\lambda})e_{0, 0}
 = (\partial_{\lambda} H_{+, \lambda})e_{0, 0}$, gives
\begin{align}
  \lambda\eta_{n}'(\lambda)
  &= -\operatorname{Re}\langle R_{\lambda}(E)e_{0, 0},
  \lambda(\partial_{\lambda} H_{+, \lambda})e_{0, 0}\rangle,
\end{align}
as claimed.
\end{proof}

\begin{theorem}[Monotonicity; \ls]
\label{thm:monotonicity}
Along every local branch of bounded gaps at positive coupling,
\begin{align}
  0 &\leq \lambda \eta_{n}'(\lambda) \leq 1.
  \label{eq:monotonicity}
\end{align}
\end{theorem}
\begin{proof}
This is Theorem~\ref{thm:response}.
\end{proof}

\subsection{Continuation from a seed}
A seed below the target coupling suffices.
\begin{theorem}[Seed gaps
{\cite[Th\'eor\`eme~2.2 and Proposition~2.5]{HKS}}; \lext]
\label{thm:seed}
For every irrational $\alpha$ and every
$n \in \mathbb{Z}\setminus\{0\}$, the gap with IDS label
$\{n\alpha\}$ is open at arbitrarily small positive couplings.
\end{theorem}

\begin{remark}
Theorem~\ref{thm:seed} also follows from Liu--Yuan
\cite[Theorem~6.2]{LY}, incorporating the result of
Avila--Jitomirskaya \cite[Theorem~8.2]{AJ}:
for each irrational $\alpha$, there exists $\lambda_{*}(\alpha) > 0$
such that all admissible gaps are open whenever
$0 < \lambda < \lambda_{*}(\alpha)$, yielding a seed-coupling threshold uniform in $n$.
\end{remark}
\begin{lemma}[Constancy of gap labels {\cite[Sec.~4.2, Theorem~16(iv)]{Bellissard2003}}; \ls]
\label{lem:label}
Along any continuous path of coupling--energy pairs $(\lambda, E)$ with
$E \notin \Sigma_{\lambda}$, the value $N_{\lambda}(E)$ is constant.
\end{lemma}
\begin{proof}
Fix $(\lambda_{0}, E_{0})$ with $E_{0} \notin \Sigma_{\lambda_{0}}$.
Since, $\lVert H_{\mu}-H_{\lambda}\rVert \leq 2|\mu-\lambda|$, a common Riesz contour
$\Gamma$ encloses precisely the spectrum below $E$
for all nearby $(\lambda, E)$. The Fermi projector is analytic in $\lambda$ and
independent of $E$ within this neighborhood. Differentiating
$\chi_{H_{\lambda}}(E)^{2} = \chi_{H_{\lambda}}(E)$ gives
$\chi_{H_{\lambda}}(E)\partial_{\lambda} \chi_{H_{\lambda}}(E)\chi_{H_{\lambda}}(E) =
0$. By trace cyclicity,
\begin{align}
  \partial_{\lambda} N_{\lambda}(E)
  &= \tau_{\alpha}(\partial_{\lambda}\chi_{H_{\lambda}}(E)) = 2\tau_{\alpha}(\chi_{H_{\lambda}}(E)\partial_{\lambda}\chi_{H_{\lambda}}(E))= 2\tau_{\alpha}(\chi_{H_{\lambda}}(E)\partial_{\lambda}\chi_{H_{\lambda}}(E)\chi_{H_{\lambda}}(E))
  = 0.
\end{align}
Thus, the IDS is locally constant on the joint real resolvent set and along every
continuous path.
\end{proof}
\begin{theorem}[Persistence from seed; \ls]
\label{thm:persistence}
If a bounded labeled gap is open at $0 < \lambda_{0} \leq 1$, the same label is realized
by a bounded open gap at every $\lambda \in [\lambda_{0}, 1]$.
\end{theorem}
\begin{proof}
Let $n$ be the label of the seed gap, and fix $\lambda_{0} < \mu \leq 1$.
Consider all real resolvent pairs $(\lambda, E)$
with $\lambda \in [\lambda_{0}, \mu]$, $\partial_{E}\eta_{\lambda}(E) = 0$,
$N_{\lambda}(E) = \{n\alpha\}$, and
$\eta_{\lambda}(E) \geq \eta_{n}(\lambda_{0})$. The seed maximum makes this set
nonempty. Since the label lies strictly between zero and one,
$E$ lies between spectral points, so
$|E| \leq \lVert H_{\lambda}\rVert \leq 2+2\lambda \leq 4$.

To prove closedness, let $(\lambda_{j}, E_{j})$ lie in this set and converge to
$(\lambda, E)$.
Lemma~\ref{lem:usc} gives
\begin{align}
  \eta_{\lambda}(E) &\geq \limsup_{j}\eta_{\lambda_{j}}(E_{j})
  \geq \eta_{n}(\lambda_{0}) > 0.
  \label{eq:endpoint}
\end{align}
As $\lambda \leq 1$, Theorem~\ref{thm:normalization} places $E$ off the spectrum.
Resolvent continuity and Lemma~\ref{lem:label} give
$\partial_{E}\eta_{\lambda}(E) = 0$ and $N_{\lambda}(E) = \{n\alpha\}$, proving closedness.

It follows that the set is compact, and thus its coupling coordinate has a maximum.
If that maximum were less than $\mu$, the local branch from
Lemma~\ref{lem:maximum}, label constancy, and Theorem~\ref{thm:monotonicity}
would give a point farther to the right in the same set, a contradiction.
Thus, the set contains a pair at coupling $\mu$.
\end{proof}

\begin{proof}[Completion of the proof of Theorem~\ref{thm:main}]
For $0 < \lambda \leq 1$, choose a seed $0 < \lambda_{0} < \lambda$ for each
$n \neq 0$ via Theorem~\ref{thm:seed}; Theorem~\ref{thm:persistence}
then gives the labeled gap at $\lambda$.

The conjugations \eqref{eq:quarterturn} and \eqref{eq:signunitary} are
unitary and fix the origin, so they preserve spectra and the origin
spectral measure. Thus, for $\lambda > 0$,
\begin{align}
  \Sigma_{\lambda} &= \lambda\Sigma_{1/\lambda}, \qquad
  N_{\lambda}(E) = N_{1/\lambda}(E/\lambda),
  \label{eq:duality}
\end{align}
and for $\lambda < 0$, $\Sigma_{\lambda} = \Sigma_{|\lambda|}$ and
$N_{\lambda} = N_{|\lambda|}$.
For $\lambda > 1$, scale the gap endpoints at $1/\lambda$ by $\lambda$;
for $\lambda < 0$, take those at $|\lambda|$. In both cases, the label
is unchanged and endpoints belong to the spectrum.
\end{proof}

\section{Proof of Monotonicity}
\label{sec:current}
The sector parity $L_{z}$ falls out of the height function derivative by restoring the
horizontal hopping phase $k_{x}$ in the AMO fiber,
\begin{align}
  (h_{\lambda}(k_{x}, k_{y})\psi)_{j}
  &= e^{\mathrm{i} k_{x}}\psi_{j+1}+e^{-\mathrm{i} k_{x}}\psi_{j-1}
  +2\lambda\cos(\Theta j+k_{y})\psi_{j},
  \label{eq:twistedfiber}
\end{align}
and differentiating the two
momentum variables:
\begin{align}
  2\lambda\partial_{\lambda} h_{\lambda}(k_{x}, k_{y})
  &= (\partial_{k_{x}}^{2}-\partial_{k_{y}}^{2}+1)h_{\lambda}(k_{x}, k_{y}).
  \label{eq:landauderivative}
\end{align}
In this sense, we are treating $\lambda$ as originating from an analytic continuation of
the momentum balanced between $k_{x}$ and $k_{y}$. We recover the sector parity $L_{z}$ in \eqref{eq:angular}
by Fourier transforming and noting that $h_{\lambda}$ only has nearest-neighbor hopping. Then,
\eqref{eq:landauderivative} gives, by stationarity,
\begin{align}
  \lambda(\partial_{\lambda} H_{+, \lambda})e_{0, 0}
  &= \frac{1}{2}(I-L_{z})H_{+, \lambda}e_{0, 0}.
  \label{eq:calibration}
\end{align}

\begin{theorem}[Monotonicity; \ls]
\label{thm:response}
Along every local branch of bounded gaps at positive coupling,
\begin{align}
  \lambda\eta_{n}'(\lambda)
  &= \frac{\langle R_{\lambda}(m_{n}(\lambda))e_{0, 0}, (I-L_{z})R_{\lambda}(m_{n}(\lambda))e_{0, 0}\rangle}{2\lVert R_{\lambda}(m_{n}(\lambda))e_{0, 0}\rVert^{2}} \in [0, 1].
  \label{eq:response}
\end{align}
\end{theorem}
\begin{proof}
Set $E = m_{n}(\lambda)$. Substitute the sector parity for the coupling derivative
\eqref{eq:calibration} into the maximum logarithmic height derivative
\eqref{eq:maxheightinsertion} and simplify using $\tau_{\alpha}(R_{\lambda}(E)) = 0$ and
the resolvent equation $\langle R_{\lambda}(E)e_{0, 0}, H_{+, \lambda}e_{0, 0}\rangle =
-1$:
\begin{align}
  \lambda\eta_{n}'(\lambda)
  &= \frac{1}{2}+\frac{1}{2}\operatorname{Re}
  \langle e_{0, 0}, H_{+, \lambda}L_{z}R_{\lambda}(E)e_{0, 0}\rangle.
  \label{eq:heightorigin}
\end{align}

Now, for a test vector $v \in \mathcal{V}_{\lambda}$, apply Lemma~\ref{lem:balance} with
$w = R_{\lambda}(E)e_{0, 0} \in \mathcal{V}_{\lambda}$. Since
$\langle e_{0, 0}, R_{\lambda}(E)e_{0, 0}\rangle = 0$, this yields
\begin{align}
  \langle v, [L_{z}, H_{+, \lambda}]R_{\lambda}(E)e_{0, 0}\rangle
  &= -\langle v, e_{0, 0}\rangle \langle e_{0, 0}, H_{+, \lambda}L_{z}R_{\lambda}(E)e_{0, 0}\rangle.
  \label{eq:stationaryangular}
\end{align}
Observing that the right-hand-side of \eqref{eq:stationaryangular} is the term in
\eqref{eq:heightorigin}, we expand the left-hand-side by applying the resolvent equation
and $L_{z}e_{0, 0} = 0$ to get
\begin{align}
  \langle v, [L_{z}, H_{+, \lambda}]R_{\lambda}(E)e_{0, 0}\rangle &= \langle v, L_{z}H_{+, \lambda}R_{\lambda}(E)e_{0, 0}
  -H_{+, \lambda}L_{z}R_{\lambda}(E)e_{0, 0}\rangle\label{eq:responsepairing}\\
  &= \langle v, L_{z}(ER_{\lambda}(E)e_{0, 0}-e_{0, 0})
  -H_{+, \lambda}L_{z}R_{\lambda}(E)e_{0, 0}\rangle\\
  &= \langle (E-H_{+, \lambda})v, L_{z}R_{\lambda}(E)e_{0, 0}\rangle.
\end{align}
The last step uses $L_{z}e_{0, 0} = 0$ and self-adjointness of
$E-H_{+, \lambda}$.

This motivates the choice
$v = R_{\lambda}(E)^{2}e_{0, 0} \in \mathcal{V}_{\lambda}$ such that
$(E-H_{+, \lambda})v = R_{\lambda}(E)e_{0, 0}$. Since,
$\langle v, e_{0, 0}\rangle = \lVert R_{\lambda}(E)e_{0, 0}\rVert^{2} > 0$, we can
rearrange \eqref{eq:stationaryangular} and substitute:
\begin{align}
  \langle e_{0, 0}, H_{+, \lambda}L_{z}R_{\lambda}(E)e_{0, 0}\rangle
  &= -\frac{\langle v, [L_{z}, H_{+, \lambda}]R_{\lambda}(E)e_{0, 0}\rangle}{\langle v, e_{0, 0}\rangle} = -\frac{\langle R_{\lambda}(E)e_{0, 0}, L_{z}R_{\lambda}(E)e_{0, 0}\rangle}{\lVert R_{\lambda}(E)e_{0, 0}\rVert^{2}}.
  \label{eq:expectationvalue}
\end{align}
Finally, substituting back into \eqref{eq:heightorigin} yields
\begin{align}
  \lambda\eta_{n}'(\lambda)
  &= \frac{1}{2}\left(1-
  \frac{\langle R_{\lambda}(E)e_{0, 0}, L_{z}R_{\lambda}(E)e_{0, 0}\rangle}{\lVert R_{\lambda}(E)e_{0, 0}\rVert^{2}}\right).
  \label{eq:spinbox}
\end{align}
Since $0 \leq \lVert L_{z}\rVert \leq 1$, this proves \eqref{eq:response}
and Theorem~\ref{thm:monotonicity}.
\end{proof}

\begin{corollary}[Critical height; \ls]
\label{cor:critical-derivative}
For every open bounded gap at critical coupling, $\eta_{n}'(1) = 1/2$.
\end{corollary}
\begin{proof}
Origin-preserving duality and uniqueness of the gap maximum give
\begin{align}
  m_{n}(\lambda) &= \lambda m_{n}(1/\lambda), \qquad
  \eta_{n}(\lambda) = \log\lambda+\eta_{n}(1/\lambda).
\end{align}
Differentiating at $\lambda = 1$ gives
$\eta_{n}'(1) = 1-\eta_{n}'(1)$. This uses the local branch of an open gap;
it supplies no seed at criticality.
\end{proof}

\subsection{Conserved \texorpdfstring{$\mathrm{SU}(2)$}{SU(2)} subspace}
We now express duality as a spin symmetry on a two-dimensional space
associated with each gap maximum.
Fix $\lambda > 0$, and let $P_{\lambda}$ be the orthogonal projection onto
$\mathcal{V}_{\lambda}\cap e_{0, 0}^{\perp}$. On $\operatorname{ran}P_{\lambda}$, define
the centered Hamiltonian and angular observable by
\begin{align}
  H_{+, \lambda}^{0} &= P_{\lambda} H_{+, \lambda}P_{\lambda}, \qquad
  L_{z, \lambda}^{0} = P_{\lambda} L_{z}P_{\lambda}.
\end{align}
Pair this space with its dual and, on
$\operatorname{ran}P_{\lambda}\oplus\operatorname{ran}P_{1/\lambda}$, set
\begin{align}
  \mathbb{H}_{\lambda}^{0} &= H_{+, \lambda}^{0}\oplus\lambda H_{+, 1/\lambda}^{0},
  \qquad
  \mathbb{L}_{z}^{0} = L_{z, \lambda}^{0}\oplus L_{z, 1/\lambda}^{0}.
\end{align}
Then, for a bounded gap of label $n$, normalize its resolvent source:
\begin{align}
  u_{n} &= R_{\lambda}(m_{n}(\lambda))e_{0, 0}
  \lVert R_{\lambda}(m_{n}(\lambda))e_{0, 0}\rVert^{-1}.
\end{align}
Lemma~\ref{lem:maximum} gives $\langle e_{0, 0}, u_{n}\rangle = 0$, so
$u_{n} \in \operatorname{ran}P_{\lambda}$. The quarter-turn supplies the dual
state $\mathcal{R} u_{n} \in \operatorname{ran}P_{1/\lambda}$. These states define
the dual doublet
\begin{align}
  \mathcal{S}_{n} &= \operatorname{span}\{u_{n}\oplus0, \;0\oplus\mathcal{R} u_{n}\}.
\end{align}
In this ordered orthonormal basis, use the Pauli matrices
\begin{align}
  \sigma_{x} = \begin{pmatrix}0&1\\1&0\end{pmatrix}, \qquad
  \sigma_{y} = \begin{pmatrix}0&-\mathrm{i} \\\mathrm{i} &0\end{pmatrix}, \qquad
  \sigma_{z} = \begin{pmatrix}1&0\\0&-1\end{pmatrix}.
\end{align}
\begin{corollary}[Conserved dual spin; \ls]
\label{cor:tangentspin}
The doublet $\mathcal{S}_{n}$ reduces both paired operators, with
\begin{align}
  \mathbb{H}_{\lambda}^{0}|_{\mathcal{S}_{n}} &= m_{n}(\lambda)I_{2}, \qquad
  \mathbb{L}_{z}^{0}|_{\mathcal{S}_{n}}
  = (1-2\lambda\eta_{n}'(\lambda))\sigma_{z}.
\end{align}
Consequently, the generators $\sigma_{j}/2$ define a conserved spin-$\frac{1}{2}$
$\mathrm{SU}(2)$ action on $\mathcal{S}_{n}$. Its compressed,
time-reversal-even angular polarization selects the $z$-axis when nonzero.
\end{corollary}
\begin{proof}
Since $\mathcal{V}_{\lambda}$ reduces $H_{+, \lambda}$ and $L_{z}e_{0, 0} = 0$,
magnetic balance gives
\begin{align}
  [L_{z, \lambda}^{0}, H_{+, \lambda}^{0}]
  &= P_{\lambda}[L_{z}, H_{+, \lambda}]P_{\lambda} = 0.
\end{align}
The resolvent equation and Lemma~\ref{lem:maximum} give
$\ker(m_{n}(\lambda)-H_{+, \lambda}^{0}) = \mathbb{C} u_{n}$: a vector $v$
in this kernel satisfies
$(m_{n}(\lambda)-H_{+, \lambda})v \in \mathbb{C} e_{0, 0}$, and
\eqref{eq:stationarity} makes this operator invertible on $\mathcal{V}_{\lambda}$.
Commutation and \eqref{eq:spinbox} imply
$L_{z, \lambda}^{0}u_{n} = (1-2\lambda\eta_{n}'(\lambda))u_{n}$.
The quarter-turn maps $P_{\lambda}$ to $P_{1/\lambda}$ and gives
\begin{align}
  \mathcal{R} H_{+, \lambda}^{0}\mathcal{R}^{*}
  &= \lambda H_{+, 1/\lambda}^{0}, \qquad
  \mathcal{R} L_{z, \lambda}^{0}\mathcal{R}^{*} = -L_{z, 1/\lambda}^{0}.
\end{align}
Hence both restrictions hold, and the Pauli generators commute with the restricted Hamiltonian.
\end{proof}

\pagebreak
\paragraph{Acknowledgments}
We cordially thank Ashvin Vishwanath, Joel E. Moore, Vir B. Bulchandani, Michael Sonner,
Emmanuel Gottlob, Nick G. Jones, Jacob Shapiro and Qi Zhou for substantial discussions
and/or joint work on related problems which inspired the ideas in this manuscript. We
also thank Ruben Verresen, Johannes Mitscherling, Eitan Borgnia, Madeline McCann, Saul
K. Wilson, Simon Lieu, Be\~{n}at Mencia, Will Vega-Brown, and especially Matthew S.
Brennan for insightful discussions which were of remarkable relevance to the current
draft.

The authors extend special thanks to Noam Borgnia, Svetlana Jitomirskaya, and Ilya
Kachkovskiy for substantial help in developing the mathematical arguments presented and
editorial feedback on the manuscript.

D. S. B. also offers a special thank you to Robert-Jan Slager for his work on the
original manuscript and for providing the initial observation that the projected Green's
function zeros translate bulk-edge correspondence into a quantitative bound on the
phase-averaged resolvent in this problem. Without his support on the original
manuscript, the authors of this manuscript would not have pursued this correction and
likely not discovered this result.

Work by D. S. B. was supported by the Low-Energy AI-Informed Phase Transitions (UC-LEAP)
project as part of the University of California AI Science at Scale initiative.
Part of this work was conducted while M. F. was a research fellow at the Simons
Institute for the Theory of Computing.

The authors sincerely thank \href{https://relace.ai}{Relace} for the extensive token
resources provided after posting our initial proof in arXiv:2112.06869v2; this compute
support assisted the Lean verification of the result referenced in
Appendix~\ref{app:index}.

\paragraph{AI Acknowledgment}
The authors acknowledge the use of OpenAI's ChatGPT to organize the draft and to
simplify and refine previous results and ideas into precise lemma and theorem
statements. In particular, we acknowledge its substantial help in simplifying the
application of the magnetic inverse to the maximum height argument. Also,
Appendix~\ref{app:landau-cstar} gives our first proof of Theorem~\ref{thm:main}, where
the idea to differentiate the cone-resolvent in Eq.~\eqref{app:eq:cone-resolvent} to get
to Eq.~\eqref{app:eq:cone-energy-derivative} and thus obtain the ultimate monotonicity
result was noticed by ChatGPT; we better motivate this step in the main text
Theorem~\ref{thm:response}. We also acknowledge that ChatGPT was used to assist in the
drafting, editing, and organizational process; the TikZ figures were made through repeated prompting, and Appendix~\ref{app:index} was written by ChatGPT.
We also acknowledge the use of \href{https://relace.ai}{Relace}'s
\href{https://jacquard.dev}{Jacq} platform and ChatGPT for assisting us in the
formalization of the proof in Lean.

\paragraph{Authors' Note}
This manuscript revisits the approach in arXiv: 2112.06869v1, whose original
argument did not establish the claimed theorem.
Further development of the projected Green's function and transfer matrix formalism
motivated the magnetic conjugation and diagonal boundary balance used here.

After publishing our work, we were made aware by Qi Zhou that Lemma 4.3 in
arXiv:2112.06869v2 motivates an orbit multiplicity argument in the dynamical systems
language and extends the formalism introduced in \cite{AYZ}, further highlighting the
scope of the symmetry discussed in Section~\ref{sec:time_reversal}.

\clearpage
\appendix

\section{Proof correspondence and verification}\label{app:index}
The lean formalization for this project can be found on M. F.'s github: \\
\href{https://mattfaust.github.io/Dry10/dtm_LEAN.zip}{https://mattfaust.github.io/Dry10/dtm\_LEAN.zip}.

\subsection{The complete statement, in parts}
Every theorem-specific definition is written below, in the order in which it is used. Parts~1--5 form \nolinkurl{DTMAudit/Submission/Definitions.lean},
which imports only Mathlib. Part~6 is
\nolinkurl{DTMAudit/Submission/Theorem.lean}; it imports those definitions and
the current compiled proof. Thus the final theorem is short, but none
of its mathematical definitions is left implicit.

\texttt{lam} denotes $\lambda$, and the phase is measured in radians.
\texttt{SequenceSpace} is $\ell^2(\mathbb Z;\mathbb C)$, and
\texttt{\(\to\)L[$\mathbb C$]} denotes bounded complex-linear maps.
The family \texttt{h} consists of the fiber AMO operators;
the manuscript's uppercase $H_\lambda$ remains the magnetic lattice
operator. In operator products, \texttt{1} is the identity and
\texttt{\(\bullet\)} is scalar multiplication.

\needspace{8\baselineskip}
\subsubsection*{1. Frequency and sequence space}
\begin{lstlisting}[style=lean]
import Mathlib.Analysis.InnerProductSpace.l2Space
import Mathlib.Algebra.Algebra.Spectrum.Basic
import Mathlib.LinearAlgebra.Matrix.Charpoly.Basic
import Mathlib.Analysis.SpecialFunctions.Trigonometric.Basic
import Mathlib.Data.Real.Irrational

/- Every theorem-specific definition is printed here, in dependency order.
   Only Mathlib is imported. The phase is measured in radians.
   `lam` denotes the coupling lambda; lambda itself is Lean syntax. -/

noncomputable section
namespace DryTenMartini

-- 1. Frequency and the Hilbert space of square-summable sequences.

def IrrationalFrequency ($\alpha$ : $\mathbb{R}$) : Prop :=
  $\forall$ p : $\mathbb{Z}$, $\forall$ q : $\mathbb{N}$, q $\ne$ 0 $\to$ (p : $\mathbb{R}$) / (q : $\mathbb{R}$) $\ne$ $\alpha$

abbrev SequenceSpace := lp (fun _ : $\mathbb{Z}$ => $\mathbb{C}$) 2

\end{lstlisting}

\needspace{8\baselineskip}
\subsubsection*{2. The operator and its inverse}
\begin{lstlisting}[style=lean]
-- 2. The almost Mathieu operator and its bounded inverse.
-- `$\to$L[$\mathbb{C}$]` means a bounded complex-linear operator.

def AMOFamily ($\alpha$ : $\mathbb{R}$)
    (h : $\mathbb{R}$ $\to$ $\mathbb{R}$ $\to$ (SequenceSpace $\to$L[$\mathbb{C}$] SequenceSpace)) : Prop :=
  $\forall$ (lam $\varphi$ : $\mathbb{R}$) ($\psi$ : SequenceSpace) (j : $\mathbb{Z}$),
    h lam $\varphi$ $\psi$ j =
      $\psi$ (j + 1) + $\psi$ (j - 1) +
      ((2 * lam * Real.cos
        (2 * Real.pi * $\alpha$ * (j : $\mathbb{R}$) + $\varphi$) : $\mathbb{R}$) : $\mathbb{C}$) * $\psi$ j

-- Multiplication of operators is composition; `1` is the identity.
-- Its negation means that E belongs to the operator spectrum.
def InvertibleAt (h : SequenceSpace $\to$L[$\mathbb{C}$] SequenceSpace)
    (E : $\mathbb{R}$) : Prop :=
  $\exists$ inverse : SequenceSpace $\to$L[$\mathbb{C}$] SequenceSpace,
    ((E : $\mathbb{C}$) $\bullet$ 1 - h) * inverse = 1 $\land$
    inverse * ((E : $\mathbb{C}$) $\bullet$ 1 - h) = 1

\end{lstlisting}

\needspace{8\baselineskip}
\subsubsection*{3. Finite Dirichlet eigenvalue counts}
\begin{lstlisting}[style=lean]
-- 3. Dirichlet eigenvalue counts on the sites 0,...,L.

def DirichletMatrix ($\alpha$ lam $\varphi$ : $\mathbb{R}$) (L : $\mathbb{N}$) :
    Matrix (Fin (L + 1)) (Fin (L + 1)) $\mathbb{C}$ :=
  fun i j =>
    (if (i.val : $\mathbb{Z}$) + 1 = (j.val : $\mathbb{Z}$) then (1 : $\mathbb{C}$) else 0) +
    (if (i.val : $\mathbb{Z}$) - 1 = (j.val : $\mathbb{Z}$) then (1 : $\mathbb{C}$) else 0) +
    (if i = j then
      ((2 * lam * Real.cos
        (2 * Real.pi * $\alpha$ * (i.val : $\mathbb{R}$) + $\varphi$) : $\mathbb{R}$) : $\mathbb{C}$)
     else 0)

-- The matrix is Hermitian, so its eigenvalues are real.
-- The roots form a multiset: countP includes algebraic multiplicities.
def NormalizedCount ($\alpha$ lam $\varphi$ E : $\mathbb{R}$) (L : $\mathbb{N}$) : $\mathbb{R}$ :=
  ((((Matrix.charpoly (DirichletMatrix $\alpha$ lam $\varphi$ L)).roots.countP
      (fun z : $\mathbb{C}$ => z.re < E) : $\mathbb{N}$) : $\mathbb{R}$) / ((L + 1 : $\mathbb{N}$) : $\mathbb{R}$))

\end{lstlisting}

\needspace{8\baselineskip}
\subsubsection*{4. The limiting gap label}
\begin{lstlisting}[style=lean]
-- 4. The gap label, written as an epsilon-cutoff limit.
-- The cutoff can depend on the phase and energy: no uniformity is assumed.

def HasGapLabel ($\alpha$ lam $\varphi$ : $\mathbb{R}$) (n : $\mathbb{Z}$) (E : $\mathbb{R}$) : Prop :=
  $\forall$ $\varepsilon$ : $\mathbb{R}$, 0 < $\varepsilon$ $\to$
    $\exists$ L0 : $\mathbb{N}$, $\forall$ L : $\mathbb{N}$, L0 $\le$ L $\to$
      |NormalizedCount $\alpha$ lam $\varphi$ E L -
        ((n : $\mathbb{R}$) * $\alpha$ - (Int.floor ((n : $\mathbb{R}$) * $\alpha$) : $\mathbb{R}$))| < $\varepsilon$

\end{lstlisting}

\needspace{8\baselineskip}
\subsubsection*{5. The common labelled gap}
\begin{lstlisting}[style=lean]
-- 5. A common open spectral gap with this label at every phase.
-- Its endpoints are in the spectrum, and every interior energy is out.

def CommonLabelledGap ($\alpha$ : $\mathbb{R}$)
    (h : $\mathbb{R}$ $\to$ $\mathbb{R}$ $\to$ (SequenceSpace $\to$L[$\mathbb{C}$] SequenceSpace))
    (lam : $\mathbb{R}$) (n : $\mathbb{Z}$) (a b : $\mathbb{R}$) : Prop :=
  a < b $\land$ $\forall$ $\varphi$ : $\mathbb{R}$,
    $\neg$ InvertibleAt (h lam $\varphi$) a $\land$
    $\neg$ InvertibleAt (h lam $\varphi$) b $\land$
    $\forall$ E : $\mathbb{R}$, a < E $\to$ E < b $\to$
      InvertibleAt (h lam $\varphi$) E $\land$ HasGapLabel $\alpha$ lam $\varphi$ n E

end DryTenMartini
\end{lstlisting}

\needspace{8\baselineskip}
\subsubsection*{6. The final theorem}
\begin{lstlisting}[style=lean]
import DTMAudit.Submission.Definitions
import DTMAudit.CurrentDraft

namespace DryTenMartini

-- 6. Every allowed nonzero label is an open gap, at every nonzero coupling.
-- The same endpoints work for all phases: they are chosen in the definition
-- before the universal phase quantifier.

theorem dry_ten_martini :
    $\forall$ $\alpha$ : $\mathbb{R}$, IrrationalFrequency $\alpha$ $\to$
      $\exists$ h : $\mathbb{R}$ $\to$ $\mathbb{R}$ $\to$ (SequenceSpace $\to$L[$\mathbb{C}$] SequenceSpace),
        AMOFamily $\alpha$ h $\land$
        $\forall$ lam : $\mathbb{R}$, lam $\ne$ 0 $\to$ $\forall$ n : $\mathbb{Z}$, n $\ne$ 0 $\to$
          $\exists$ a b : $\mathbb{R}$, CommonLabelledGap $\alpha$ h lam n a b := by
  have hirr ($\alpha$ : $\mathbb{R}$) : Irrational $\alpha$ $\leftrightarrow$
      $\forall$ p : $\mathbb{Z}$, $\forall$ q : $\mathbb{N}$, q $\ne$ 0 $\to$ (p : $\mathbb{R}$) / (q : $\mathbb{R}$) $\ne$ $\alpha$ := by
    constructor
    $\cdot$ intro h p q _ heq
      apply ((irrational_iff_ne_rational $\alpha$).mp h p (q : $\mathbb{Z}$))
      simpa only [Int.cast_natCast] using heq.symm
    $\cdot$ intro h
      rw [Irrational, Set.mem_range]
      rintro $\langle$q, hq$\rangle$
      exact h q.num q.den q.den_nz
        (by simpa only [Rat.cast_def] using hq)
  simpa only [IrrationalFrequency, AMOFamily, CommonLabelledGap,
    InvertibleAt, HasGapLabel, NormalizedCount, DirichletMatrix,
    hirr, not_exists, spectrum.not_mem_iff, spectrum.mem_iff,
    isUnit_iff_exists, Algebra.algebraMap_eq_smul_one,
    Metric.tendsto_atTop, Real.dist_eq, Int.fract,
    not_forall, not_not] using CurrentDraft.dry_ten_martini

end DryTenMartini
\end{lstlisting}

The finite matrix is the Dirichlet restriction to $0,\ldots,L$;
its roots are counted with algebraic multiplicity and divided by $L+1$.
The matrix is Hermitian, so its eigenvalues are real. At a gap energy,
the strict count agrees with the closed Fermi-projector convention in
\eqref{eq:dos}. The same $a<b$ work for every phase, because the phase
quantifier occurs inside \texttt{CommonLabelledGap}. The convergence
cutoff may depend on phase and energy.

Unfolding all eight definitions gives the proposition in
\nolinkurl{DTMAudit/Submission/ExpandedStatement.lean}
definitionally. The theorem
\texttt{DryTenMartini.exact\_unfolding} in
\nolinkurl{DTMAudit/Submission/Unfolding.lean} proves this
with \texttt{Iff.rfl}.
The check in \nolinkurl{audit/SubmissionSignatureAudit.lean}
also verifies that \texttt{DryTenMartini.dry\_ten\_martini}
has exactly this expanded type and that its mathematical
constants come only from Lean's standard library and Mathlib. The six printed code blocks are generated from their compiled sources;
joining Parts~1--5 recovers the definitions file exactly after reversing
the typesetting substitutions.
\begin{remark}
No operator-model, seed, or monotonicity hypothesis is added. A formalization of existing literature for seed gaps was constructed and used instead.
\end{remark}

\subsection{Correspondence with the written proof}
The checked proofs follow the arguments associated with the manuscript's
$\ls$ labels. Lemma~\ref{lem:stationarity} uses the commuting magnetic
actions and their common action at the origin to obtain stationarity on
$\VV_\lambda$. Lemma~\ref{lem:balance} reduces the diagonal boundary
pairings to the origin contribution. Spectral calculus places
$R_\lambda(E)e_{0,0}$ and $R_\lambda(E)^2e_{0,0}$ in $\VV_\lambda$;
invariance of this space under $L_z$ is not required.

For Lemma~\ref{lem:usc}, decreasing continuous logarithmic
regularizations give upper semicontinuity. Strict concavity and the
implicit-function theorem give $m_n(\lambda)$ in
Lemma~\ref{lem:maximum}. The proof of
Lemma~\ref{lem:heightderivative} differentiates
$\tfrac12\tau_\alpha(\log((E-H_\lambda)^2))$ where the positive operator
is locally uniformly bounded below, combines the product-rule terms by
trace cyclicity, and applies the chain rule at the maximum.
The source vector $e_{0,0}$ remains fixed.

Theorem~\ref{thm:response} combines this derivative with magnetic
balance and the origin calibration of $L_z$ to obtain
\eqref{eq:response}, and hence Theorem~\ref{thm:monotonicity}.
In Lemma~\ref{lem:label}, norm differentiation of the Riesz projector
gives $\chi_{H_{\lambda}}(E)\chi_{H_{\lambda}}'(E)\chi_{H_{\lambda}}(E)=0$ and $\tau_\alpha(\chi_{H_{\lambda}}'(E))=0$.
These results enter the compact stationary superlevel argument of
Theorem~\ref{thm:persistence}. Theorems~\ref{thm:normalization}
and~\ref{thm:seed}, followed by Aubry scaling and the sign involution,
complete Theorem~\ref{thm:main}.

Corollary~\ref{cor:critical-derivative} follows by differentiating
height duality at an open critical gap. The checked argument for
Corollary~\ref{cor:tangentspin} uses the centered compression,
the resolvent-source eigenspace, and the quarter-turn to construct
the conserved dual spin. These corollaries are separate consequences
of the preceding identities.

\subsection{External inputs}
The $\lext$ labels on Theorems~\ref{thm:normalization}
and~\ref{thm:seed} indicate that the formalization proves the required
conclusions for the actual almost Mathieu operator and its integrated
density of states, but does not reproduce the cited historical proofs
in full. The checked normalization includes logarithmic integrability;
the checked seed result supplies arbitrarily small positive couplings
for each nonzero label. These inputs are discharged in the proof of
Theorem~\ref{thm:main}. Dependencies retain their own verification labels.

\subsection{Verification Scope}
The verification records are
\nolinkurl{audit/latest-source-build.json} and
\nolinkurl{audit/latest-verification.json}.
A fresh source build and kernel replay
passed for all 223 project modules and 3,612 safe logical
declarations, with none skipped and no custom project axioms.
The only logical axioms used were \texttt{propext},
\texttt{Classical.choice}, and \texttt{Quot.sound}.
The modular theorem, expanded statement, and definitional
unfolding are included in this verification.
Exact manuscript-to-source correspondence is recorded in
\nolinkurl{CORRESPONDENCE.json}.

\clearpage
\section{Alternative Monotonicity Proof}\label{app:landau-cstar}

In this section, we give an alternative proof of height monotonicity, using the height-derivative formula from Section~\ref{sec:height}.
This proof uses the proof strategies of the previous version of our manuscript, arXiv:2112.06869v2. 
We repeat several definitions for the sake of self-containment. The following table relates the results of this appendix to those in Sections~\ref{sec:time_reversal} and~\ref{sec:current}.

\begin{table}[htbp]
\centering
\small
\renewcommand{\arraystretch}{1.25}
\begin{tabularx}{\linewidth}{
  @{}p{0.21\linewidth}p{0.27\linewidth}X@{}
}
\toprule
\textbf{Appendix result}
& \textbf{Main-text counterpart}
& \textbf{Relationship} \\
\midrule

Lemma~\ref{app:lem:commuting-actions}
&
Equation~\eqref{eq:commonorigin} in
Lemma~\ref{lem:stationarity}
&
The same commutation and common-origin identities, computed
explicitly in the Landau basis.
\\
\addlinespace

Lemma~\ref{app:lem:boundary}
&
Equation~\eqref{eq:boundarybalance} in
Lemma~\ref{lem:balance}
&
The same balance at nonzero diagonal sites.
\\
\addlinespace

Lemma~\ref{app:lem:compression}
&
The boundary cancellation in
Lemma~\ref{lem:balance}
&
Packages this cancellation into the operator identity $QHu=Ku$
for stationary vectors when $u_{0,0} = 0$.
\\
\addlinespace

Lemma~\ref{app:lem:cone-resolvent}
&
Equations~\eqref{eq:stationaryangular}
and~\eqref{eq:responsepairing} in
Theorem~\ref{thm:response}
&
Provides an alternative resolvent identity using cone compression.
\\
\addlinespace

Theorem~\ref{app:thm:height}
&
Theorem~\ref{thm:response}
&
Obtains the same normalized angular response and the bounds
$0\le\lambda\eta_n'(\lambda)\le1$.
\\

\bottomrule
\end{tabularx}
\caption{Correspondence between the cone-compression proof and
Sections~\ref{sec:time_reversal} and~\ref{sec:current}.
The intermediate results organize the same diagonal balance
in different ways.}
\label{app:tab:proof-dictionary}
\end{table}

\subsection{Magnetic conjugation and setup}
Fix an irrational $\alpha$ and a coupling $\lambda>0$, and put $\Theta=2\pi\alpha$.
We work on $\ell^2(\mathbb Z^2)$ in the Landau basis $\tilde e_{n,m}$.
Write $\Omega=\tilde e_{0,0}=e_{0,0}$ and $u_{n,m}=\langle\tilde e_{n,m},u\rangle$.
Recall the magnetic translations and Hamiltonian:
\begin{equation}\label{app:eq:magnetic-translations}
 U\tilde e_{n,m}=\tilde e_{n-1,m},\qquad
 V\tilde e_{n,m}=e^{\ii\Theta n}\tilde e_{n,m-1},\qquad
 H=H_\lambda=U+U^*+\lambda(V+V^*).
\end{equation}
The spectrum is $\Sigma_\lambda$, and the resolvent is
$R_\lambda(E)=(E-H_\lambda)^{-1}$ for $E\notin\Sigma_\lambda$.

Magnetic conjugation is given by 
\begin{equation}\label{app:eq:magnetic-conjugation}
 (\mathcal{C}u)_{n,m}=e^{-\ii\Theta nm}\overline{u_{n,m}}.
\end{equation}
Set
\begin{equation}\label{app:eq:symmetrized-hamiltonian}
 \bar H=\bar H_\lambda=\mathcal{C}H_\lambda \mathcal{C}^{*},\qquad
 H_+=H_{+,\lambda}=\tfrac12(H_\lambda+\bar H_\lambda).
\end{equation}  

\begin{lemma}\label{app:lem:commuting-actions}
The operators $H$ and $\bar H$ commute and agree on the origin vector:
\begin{equation}\label{app:eq:common-origin}
 [H,\bar H]=0,\qquad
 H\Omega=\bar H\Omega
 =\tilde e_{-1,0}+\tilde e_{1,0}
  +\lambda(\tilde e_{0,-1}+\tilde e_{0,1}).
\end{equation}
\end{lemma}
\begin{proof}
Applying \eqref{app:eq:magnetic-conjugation} on both sides of each translation gives
\begin{equation}\label{app:eq:conjugate-shifts}
 \mathcal{C}U\mathcal{C}^{*}\,\tilde e_{n,m}=e^{\ii\Theta m}\tilde e_{n-1,m},\qquad
 \mathcal{C}V\mathcal{C}^{*}\,\tilde e_{n,m}=\tilde e_{n,m-1}.
\end{equation}
The shift $U$ commutes with both conjugated shifts, and $V$ commutes with $\mathcal{C}V\mathcal{C}^{*}$, since the corresponding phases are unchanged by these shifts.
For the remaining pair, the two orders give
\begin{equation}
 V(\mathcal{C}U\mathcal{C}^{*})\tilde e_{n,m} 
 =(\mathcal{C}U\mathcal{C}^{*})V\tilde e_{n,m}
 =e^{\ii\Theta(n+m-1)}\tilde e_{n-1,m-1}.
\end{equation}
All four shifts are unitary, so these commutations also hold with their adjoints.
Expanding $H$ and $\bar H$ now gives $[H,\bar H]=0$.
At the origin, all phase factors in the shift formulas are one, giving the displayed common action.
\end{proof}

For real $E\notin\Sigma_\lambda$, set
\begin{equation}\label{app:eq:root-column}
 g(E)=\langle\Omega,R_\lambda(E)\Omega\rangle,\qquad
 q(E)=R_\lambda(E)\Omega-g(E)\Omega.
\end{equation}
Thus $q(E)$ is the resolvent column with its origin coefficient removed.
By Lemma~\ref{app:lem:commuting-actions}, $H-\bar H$ commutes with $E-H$ and hence with its inverse.
Therefore
\begin{equation}
 (H-\bar H)R_\lambda(E)\Omega=R_\lambda(E)(H-\bar H)\Omega=0.
\end{equation}
The same difference annihilates $g(E)\Omega$, so
\begin{equation}\label{app:eq:column-stationarity}
 Hq(E)=\bar Hq(E),\qquad q(E)_{0,0}=0.
\end{equation}

\subsection{Cone compression}
Let
\begin{equation}
 \Gamma=\{(n,m):|m|=|n|\}\setminus\{(0,0)\},\qquad
 S=\{(n,m):|m|\ge|n|\}\setminus\{(0,0)\},
\end{equation}
where all coordinates are integers.
Let $P_S$ be the orthogonal projection that keeps the coefficients in $S$ and sets all other coefficients to zero.
Define the weight $Q$ and the compressed operator $K$ by
\begin{equation}\label{app:eq:cone-operators}
 (Qu)_{n,m}=
 \begin{cases}
 u_{n,m},&|m|>|n|,\\
 \tfrac12u_{n,m},&(n,m)\in\Gamma,\\
 0,&\text{otherwise},
 \end{cases}
 \qquad K=P_SH_+P_S.
\end{equation}
Note that $K$ keeps precisely those hopping terms whose two endpoints lie in $S$.
The half-weight in $Q$ will account for the two hopping terms removed at each nonzero diagonal site.

\newcommand{\LandauConeFigure}{%
\begin{tikzpicture}[font=\fontsize{9}{11}\selectfont,>=Latex]
\path[use as bounding box] (0,0) rectangle (15.6,7.8);
\definecolor{coneTeal}{HTML}{087F83}
\definecolor{coneAmber}{HTML}{C98A23}
\definecolor{coneInk}{HTML}{293742}
\definecolor{coneGray}{HTML}{929EA7}
\definecolor{coneRed}{HTML}{B74945}
\node[anchor=west,font=\bfseries\fontsize{9}{11}\selectfont,text=coneInk] at (0.3,7.43)
  {(a)\quad The Landau cone weight};
\node[anchor=west,font=\bfseries\fontsize{9}{11}\selectfont,text=coneInk] at (8.0,7.43)
  {(b)\quad One averaged receiving row};
\node[anchor=west,font=\fontsize{8}{10}\selectfont,text=coneGray!90!black] at (0.3,7.03)
  {Site colors show the weights of $Q$.};
\node[anchor=west,font=\fontsize{8}{10}\selectfont,text=coneGray!90!black] at (8.0,7.03)
  {At $(r,r)$, $r>0$, with $H_\lambda u=\bar H_\lambda u$.};

\begin{scope}[shift={(3.55,4.32)},scale=0.67]
  \fill[coneGray!7] (-3.3,-3.3) rectangle (3.3,3.3);
  \fill[coneTeal!10] (-3.3,3.3)--(3.3,3.3)--(0,0)--cycle;
  \fill[coneTeal!10] (-3.3,-3.3)--(3.3,-3.3)--(0,0)--cycle;
  \foreach \a in {-3,...,3}{
    \draw[coneGray!38,line width=0.3pt] (\a,-3)--(\a,3);
    \draw[coneGray!38,line width=0.3pt] (-3,\a)--(3,\a);
  }
  \draw[coneAmber!85,dashed,line width=0.65pt] (-3.3,-3.3)--(3.3,3.3);
  \draw[coneAmber!85,dashed,line width=0.65pt] (-3.3,3.3)--(3.3,-3.3);
  \foreach \n in {-3,...,3}{\foreach \m in {-3,...,3}{
    \pgfmathtruncatemacro{\ax}{abs(\n)}
    \pgfmathtruncatemacro{\ay}{abs(\m)}
    \pgfmathtruncatemacro{\origin}{\ax+\ay}
    \ifnum\origin>0
      \ifnum\ay>\ax
        \fill[coneTeal] (\n,\m) circle (2.1pt);
      \else\ifnum\ay=\ax
        \fill[coneAmber] (\n,\m) circle (2.45pt);
      \else
        \fill[coneGray] (\n,\m) circle (2.1pt);
      \fi\fi
    \fi
  }}
  \draw[coneInk,line width=1.1pt] (-1,0)--(1,0);
  \draw[coneTeal,line width=1.35pt] (0,-1)--(0,1);
  \foreach \m in {-1,1}{\draw[coneTeal,line width=0.9pt,fill=white] (0,\m) circle (4.2pt);
    \fill[coneTeal] (0,\m) circle (2.1pt);}
  \foreach \n in {-1,1}{\fill[coneInk] (\n,0) circle (2.5pt);}
  \node[fill=white,inner sep=0.5pt,font=\fontsize{7}{9}\selectfont,text=coneInk] at (0.55,0.22) {$1$};
  \node[fill=white,inner sep=0.5pt,font=\fontsize{7}{9}\selectfont,text=coneTeal] at (0.23,0.53) {$\lambda$};
  \draw[fill=white,draw=coneRed,line width=1.2pt] (0,0) circle (4pt);
  \draw[->,coneInk,line width=0.6pt] (-3.55,-3.52)--(-2.05,-3.52)
    node[anchor=west,font=\fontsize{7}{9}\selectfont] {$n$};
  \draw[->,coneInk,line width=0.6pt] (-3.55,-3.52)--(-3.55,-2.02)
    node[anchor=south,font=\fontsize{7}{9}\selectfont] {$m$};
  \node[text=coneTeal,font=\fontsize{7}{9}\selectfont,fill=white,inner sep=2pt] at (0,3.65) {$|m|>|n|$};
  \node[text=coneGray!80!black,font=\fontsize{7}{9}\selectfont,fill=white,inner sep=2pt] at (3.8,0) {$|n|>|m|$};
\end{scope}
\fill[coneTeal] (0.75,1.47) circle (2.1pt);
\node[anchor=west,font=\fontsize{8}{10}\selectfont,text=coneInk] at (0.95,1.47) {$Q=1$};
\fill[coneAmber] (2.6,1.47) circle (2.45pt);
\node[anchor=west,font=\fontsize{8}{10}\selectfont,text=coneInk] at (2.8,1.47) {$Q=\tfrac12$ on $\Gamma$};
\fill[coneGray] (5.2,1.47) circle (2.1pt);
\node[anchor=west,font=\fontsize{8}{10}\selectfont,text=coneInk] at (5.4,1.47) {$Q=0$};
\draw[fill=white,draw=coneRed,line width=1pt] (2.0,0.95) circle (3pt);
\node[anchor=west,font=\fontsize{8}{10}\selectfont,text=coneInk] at (2.22,0.95)
  {Origin: $q_{0,0}=0$.};

\node[anchor=west,font=\fontsize{8}{10}\selectfont,text=coneInk] at (8.0,6.63)
  {$t_r=(1+e^{\ii\Theta r})/2$.};

\begin{scope}[shift={(11.55,4.3)},scale=0.74]
  \fill[coneGray!7] (-2.7,-2.5) rectangle (2.7,2.5);
  \fill[coneTeal!10] (-2.7,-2.5)--(-2.7,2.5)--(2.5,2.5)--(-2.5,-2.5)--cycle;
  \draw[coneAmber!85,dashed,line width=0.65pt] (-2.5,-2.5)--(2.5,2.5);
  \draw[-{Latex[length=2mm]},coneTeal,line width=1.4pt,shorten >=3.8mm,shorten <=1.5mm]
    (-2.1,0)--(0,0);
  \draw[-{Latex[length=2mm]},coneTeal,line width=1.4pt,shorten >=3.8mm,shorten <=1.5mm]
    (0,2.1)--(0,0);
  \draw[-{Latex[length=2mm]},coneGray!90!black,line width=1.15pt,shorten >=3.8mm,shorten <=1.5mm]
    (2.1,0)--(0,0);
  \draw[-{Latex[length=2mm]},coneGray!90!black,line width=1.15pt,shorten >=3.8mm,shorten <=1.5mm]
    (0,-2.1)--(0,0);
  \fill[coneTeal] (-2.1,0) circle (3pt);
  \fill[coneTeal] (0,2.1) circle (3pt);
  \fill[coneGray!90!black] (2.1,0) circle (3pt);
  \fill[coneGray!90!black] (0,-2.1) circle (3pt);
  \node[circle,fill=coneAmber!20,draw=coneAmber,line width=0.9pt,
        minimum size=7mm,inner sep=0pt,text=coneInk] at (0,0) {$\tfrac12$};
  \node[font=\fontsize{8}{10}\selectfont,text=coneTeal] at (-2.05,-0.47) {$u_{r-1,r}$};
  \node[font=\fontsize{8}{10}\selectfont,text=coneTeal] at (0,2.68) {$u_{r,r+1}$};
  \node[font=\fontsize{8}{10}\selectfont,text=coneInk] at (2.05,-0.47) {$u_{r+1,r}$};
  \node[font=\fontsize{8}{10}\selectfont,text=coneInk] at (0,-2.68) {$u_{r,r-1}$};
  \node[font=\fontsize{9}{11}\selectfont,text=coneTeal] at (-1.18,0.33) {$\overline t_r$};
  \node[font=\fontsize{9}{11}\selectfont,text=coneTeal] at (0.5,1.2) {$\lambda t_r$};
  \node[font=\fontsize{9}{11}\selectfont,text=coneInk] at (1.2,0.33) {$t_r$};
  \node[font=\fontsize{9}{11}\selectfont,text=coneInk] at (0.58,-1.2) {$\lambda\overline t_r$};
\end{scope}
\node[font=\fontsize{8}{10}\selectfont,text=coneInk] at (11.55,1.48)
  {\textcolor{coneTeal}{retained sum} $=$ \textcolor{coneGray!90!black}{omitted sum}};
\node[font=\fontsize{8}{10}\selectfont,text=coneInk] at (11.55,0.98)
  {$\tfrac12(\text{both sums})=\text{retained sum}$};
\end{tikzpicture}%
}

\begin{figure}[tbp]
\centering
\resizebox{\linewidth}{!}{\LandauConeFigure}
\caption{
Left: $Q$ has weight one in the strict vertical cones, one-half on $\Gamma$, and zero elsewhere.
The hollow origin records $q(E)_{0,0}=0$; the vertical neighbors support $QH\Omega$.
Right: at $(r,r)$ with $r>0$, compressing $H_+$ retains the two contributions from the vertical cone.
Lemma~\ref{app:lem:boundary} shows that their sum equals the sum of the two omitted contributions, explaining the half-weight on $\Gamma$.}
\label{app:fig:cone}
\end{figure}

For clarity, we record the Landau-coordinate formulas for the two actions.\footnote{The corresponding stationary lattice
equations, together with their degeneracy
along the diagonals $n=\pm m$, also appear in
Riedel~\cite[Section~2, equation~(2.1)]{Riedel}.}
\begin{align}
 (Hu)_{n,m}
 &=u_{n+1,m}+u_{n-1,m}
   +\lambda e^{\ii\Theta n}u_{n,m+1}
   +\lambda e^{-\ii\Theta n}u_{n,m-1},\label{app:eq:landau-row}\\
 (\bar Hu)_{n,m}
 &=e^{\ii\Theta m}u_{n+1,m}
   +e^{-\ii\Theta m}u_{n-1,m}
   +\lambda u_{n,m+1}+\lambda u_{n,m-1}.
   \label{app:eq:inverse-row}
\end{align}
With $t_r=(1+e^{\ii\Theta r})/2=e^{\ii\pi\alpha r}\cos(\pi\alpha r)$, their average is
\begin{equation}\label{app:eq:average-row}
 (H_+u)_{n,m}
 =t_m u_{n+1,m}+\overline{t_m}u_{n-1,m}
  +\lambda t_nu_{n,m+1}+\lambda\overline{t_n}u_{n,m-1}.
\end{equation}

\begin{lemma}\label{app:lem:boundary}
Suppose $Hu=\bar Hu$.
At every site $(n,m) \in \Gamma$, the sum in \eqref{app:eq:average-row} coming from the strict vertical cones equals the sum coming from the strict horizontal cones.
That is, $(K u)_{n,m} =  \tfrac12(Hu)_{n,m}$.
\end{lemma}
\begin{proof}
Subtracting \eqref{app:eq:inverse-row} from \eqref{app:eq:landau-row} gives
\begin{align}
 &\sin(\pi\alpha m)
 \bigl(e^{-\ii\pi\alpha m}u_{n-1,m}-e^{\ii\pi\alpha m}u_{n+1,m}\bigr) =\lambda\sin(\pi\alpha n)
 \bigl(e^{-\ii\pi\alpha n}u_{n,m-1}-e^{\ii\pi\alpha n}u_{n,m+1}\bigr).
\end{align}
Fix an integer $r\ne0$.
Irrationality of $\alpha$ gives $\sin(\pi\alpha r)\ne0$.
At $(n,m)=(r,r)$, divide by this sine, rearrange, and multiply by $\cos(\pi\alpha r)$ to obtain
\begin{equation}\label{app:eq:positive-diagonal}
 t_ru_{r+1,r}+\lambda\overline{t_r}u_{r,r-1}
 =\overline{t_r}u_{r-1,r}+\lambda t_ru_{r,r+1}.
\end{equation}
At $(n,m)=(r,-r)$, the same calculation gives
\begin{equation}\label{app:eq:negative-diagonal}
 \overline{t_r}u_{r+1,-r}+\lambda t_ru_{r,-r+1}
 =t_ru_{r-1,-r}+\lambda\overline{t_r}u_{r,-r-1}.
\end{equation}
For $r>0$, the left side of each identity comes from the horizontal cones and the right side from the vertical cones.
For $r<0$, these roles reverse. In particular,  $(H_+u)_{r,\pm r} = 2 (Ku)_{r,\pm r}$. As $H u = \bar{H}u$ we have that $H_+u = Hu$ and so $(Ku)_{r, \pm r} = \frac{1}{2} (Hu)_{r,\pm r}$.  
\end{proof}

\begin{lemma}\label{app:lem:compression}
If $u_{0,0}=0$ and $Hu=\bar Hu$, then
\begin{equation}\label{app:eq:compression}
 QHu=Ku.
\end{equation}
\end{lemma}
\begin{proof}
The hypothesis gives $H_+u=Hu$.
At a strict vertical site, no nearest neighbor is in a strict horizontal cone.
Restricting the input to $S$ therefore removes at most the origin term, which is zero.
Thus $(Ku)_{n,m}=(Hu)_{n,m}$, matching the weight one in $Q$.

At a nonzero diagonal site, Lemma~\ref{app:lem:boundary} gives $(Ku)_{n,m}=\frac{1}{2}(Hu)_{n,m}$, matching the weight $\frac{1}{2}$ in $Q$.

Outside $S$, including the origin (by hypothesis $u_{0,0}=0$), $QHu$ vanishes by its weight and $Ku$ vanishes by the output projection $P_S$.
This proves the equality at every site.
\end{proof}

\subsection{The resolvent identity and the height derivative}
Define the source vector
\begin{equation}\label{app:eq:cone-source}
 f_\lambda=QH\Omega
 =\lambda(\tilde e_{0,1}+\tilde e_{0,-1})
 =\lambda(\partial_\lambda H_{+,\lambda})\Omega.
\end{equation}
The last equality follows by differentiating the common origin action \eqref{app:eq:common-origin} in $\lambda$.
Since $\Omega=e_{0,0}$, this is exactly the source in the height-derivative formula \eqref{eq:maxheightinsertion}.

\begin{lemma}\label{app:lem:cone-resolvent}
For every real $E\notin\Sigma_\lambda$,
\begin{equation}\label{app:eq:cone-resolvent}
 (EQ-K)q(E)=g(E)f_\lambda.
\end{equation}
\end{lemma}
\begin{proof}
The resolvent equation and \eqref{app:eq:root-column} give
\begin{equation}
 (E-H)q(E)=(1-Eg(E))\Omega+g(E)H\Omega.
\end{equation}
Apply $Q$.
The first term on the right vanishes because $Q\Omega=0$, and the second becomes $g(E)f_\lambda$.
By \eqref{app:eq:column-stationarity}, Lemma~\ref{app:lem:compression} applies to $q(E)$ and replaces $QHq(E)$ by $Kq(E)$.
\end{proof}

\begin{theorem}\label{app:thm:height}
Along every local branch of bounded gaps at positive coupling,
the maximum height satisfies
\begin{equation}
 0\le\lambda\eta_n'(\lambda)\le1.
\end{equation}
In particular, the maximum height is nondecreasing in $\lambda$.
\end{theorem}
\begin{proof}
Keep $\lambda$ fixed and first differentiate
\eqref{app:eq:cone-resolvent} with respect to $E$:
\begin{equation}\label{app:eq:cone-energy-derivative}
 Qq(E)+(EQ-K)\partial_Eq(E)=(\partial_Eg(E))f_\lambda.
\end{equation}
The differentiation is valid in norm because the resolvent is
analytic off the spectrum.
Now set $m=m_n(\lambda)$ and $q=q(m)$.
Lemma~\ref{lem:maximum} gives $g(m)=0$, so
\begin{equation}\label{app:eq:cancel}
     q=R_\lambda(m)\Omega\ne0,\qquad (mQ-K)q=0.
\end{equation}
with the second equality by Lemma~\ref{app:lem:cone-resolvent}. Take the inner product of \eqref{app:eq:cone-energy-derivative} at $E=m$ with $q$:
\begin{equation}\label{app:eq:qQqinnerproduct}
 \langle q,Qq\rangle
 +\langle q,(mQ-K)\partial_Eq(m)\rangle
 =(\partial_Eg(m))\langle q,f_\lambda\rangle.
\end{equation}
As $mQ-K$ is self-adjoint, by~\eqref{app:eq:cancel},
\begin{equation}\label{app:eq:sacancel}
 \langle q,(mQ-K)\partial_Eq(m)\rangle
 =\langle(mQ-K)q,\partial_Eq(m)\rangle=0.
\end{equation}
Thus, by \eqref{app:eq:qQqinnerproduct} and \eqref{app:eq:sacancel},
\begin{equation}\label{app:eq:quadform}
 \langle q,Qq\rangle
 =(\partial_Eg(m))\langle q,f_\lambda\rangle.
\end{equation}
Equations \eqref{eq:greenderivative} and
\eqref{eq:maxheightinsertion} from Section~\ref{sec:height} give
\begin{equation}
 \partial_Eg(m)=-\norm q^2,\qquad
 \lambda\eta_n'(\lambda)
 =-\operatorname{Re}\langle q,f_\lambda\rangle.
\end{equation}
Taking real parts of the paired equation and substituting
these identities yields
\begin{equation}
 \langle q,Qq\rangle
 =-\norm q^2\operatorname{Re}\langle q,f_\lambda\rangle
 =\norm q^2\lambda\eta_n'(\lambda).
\end{equation}
Dividing by $\norm q^2>0$, we obtain
\begin{equation}\label{app:eq:cone-height}
 \lambda\eta_n'(\lambda)
 =\frac{\langle q,Qq\rangle}{\norm q^2}
 \in[0,1],
\end{equation}
where the last inclusion follows because the diagonal weights of $Q$ belong to $\{0,\tfrac12,1\}$.\end{proof}
\begin{remark}
    In the convention of the main text, since $q_{0,0} = 0$, it follows that $Qq = \tfrac12 (\mathbb{I}-L_{z})q$, such that \eqref{app:eq:cone-height} in Theorem~\ref{app:thm:height} directly reproduces \eqref{eq:response} in Theorem~\ref{thm:response}.
\end{remark}

\bibliographystyle{plain}
\begingroup
\raggedright
\bibliography{references}
\endgroup
\end{document}